\documentclass[12pt]{article}

\usepackage{graphicx}
\begin{document}

\hfill
{\vbox{
\hbox{UCI-TR-96-34}
\hbox{UTEXAS-HEP-96-15}
\hbox{DOE-ER-40757-084}
}}

\begin{center}
{\Large \bf Effects of Anomalous Couplings of Quarks on Prompt Photon 
Production}

\vspace{0.1in}

Kingman Cheung\footnote{Electronic mail: {\tt cheung@utpapa.ph.utexas.edu}}

{\it  Center for Particle Physics, University of 
Texas, Austin, Texas 78712} 

Dennis Silverman\footnote{Electronic mail: {\tt djsilver@uci.edu}}

{\it Department of Physics \& Astronomy, University of 
California, Irvine, CA 92697-4575}

\end{center}

\begin{abstract} 
Prompt photon production is sensitive to the anomalous couplings of
gluons to quarks, because it is mainly produced by quark-gluon
scattering.  We will examine the effects of the anomalous
chromoelectric and chromomagnetic dipole moment couplings of quarks on
prompt photon production.  Using the data collected by CDF and D0 at
the Fermilab Tevatron we put a bound on these anomalous couplings.  We
also estimate the sensitivity of various future high energy collider
experiments to these anomalous couplings
\end{abstract}

\thispagestyle{empty}

\section{Introduction}

The Standard Model (SM) has been very successful for more than thirty years.
Only recently have some deviations from the SM surfaced in the $R_b$
and $R_c$ measurements at LEP \cite{lep} 
and in the high $E_T$ inclusive jet production recorded by CDF \cite{cdf-jet}. 
Since we have no true knowledge of the structure or even the symmetry of 
the correct high energy theory,  we use the effective 
Lagrangian approach to study  low energy phenomena. 
Deviations from the SM can be studied systematically by means of an 
effective Lagrangian,  which is made up of the SM fields and obeys the
symmetry of the low energy theory.
The leading terms are simply given by the SM and consist of dimension-4
operators while the higher order terms consist of 
higher-dimension operators and are 
suppressed by powers of the scale $\Lambda$ of 
the new physics.  In other words, if the scale $\Lambda$ is much larger than
the present scale the theory is essentially the same as the SM.

Among all the dimension-5 operators, the most interesting ones
 involving quarks and gluons 
are the chromomagnetic (CMDM) and chromoelectric (CEDM) 
dipole moment couplings of quarks.  They
are given by $\sigma^{\mu\nu} T^a G^a_{\mu\nu}$ and 
$i\sigma^{\mu\nu} \gamma^5 T^a G^a_{\mu\nu}$, respectively.  
Although these couplings are zero at tree level 
within the SM, they can be induced in loop levels.    In many extensions of the
SM, they are easily nonzero at one-loop level or even tree-level, e.g.,
the multi-Higgs doublet model \cite{weinberg}. These dipole moment couplings
 are important not only because they are only suppressed by one power of 
$\Lambda$, but also because a nonzero value for the CEDM
moment is a clean signal for CP violation.  
The effects of these anomalous couplings have been studied quite extensively,
e.g., in $t\bar t$ production \cite{cheung,rizzo,haberl}, in $b\bar b$
production \cite{rizzo}, and in inclusive jet production \cite{dennis}.

The purpose of this paper is to study the effects of the anomalous 
CMDM and CEDM  of quarks on prompt photon
production.  Prompt photon production has been known to be a useful probe to
the gluon luminosity inside a hadron because they are mainly 
produced by quark-gluon scattering.
The fact that the production depends on the quark-gluon vertex also makes 
the process sensitive to the anomalous couplings of quarks to gluons.
Not only is the total cross section affected but also the differential
distributions, e.g., the transverse momentum distribution.
Both CDF\cite{cdf} and D0\cite{d0}
 have measurements on prompt photon production.  We can 
therefore use the data to constrain these CMDM and CEDM couplings.  Thus, the
bounds obtained will be the main result of the paper.  The organization is
as follows. In the next section, we shall write down the effective Lagrangian
and the formulas for the calculation.  In Sec.~III we study the effects
on the transverse momentum distribution, and obtain the results. 
In Sec.~IV we estimate the limits of these anomalous couplings that can be 
probed in the future collider experiments.  We conclude in Sec.~V.

\section{Effective Lagrangian}

The effective Lagrangian for the interactions between a quark and a gluon
that include the CEDM and CMDM form factors is 
\begin{equation}
\label{eff}
{\cal L}_{\rm eff} = g_s \bar q T^a \left[ -\gamma^\mu G_\mu^a +
\frac{\kappa}{4m_q} \sigma^{\mu\nu} G_{\mu\nu}^a - 
\frac{i \tilde{\kappa}}{4m_q} \sigma^{\mu\nu} \gamma^5 G_{\mu\nu}^a 
\right ] q \;.
\end{equation}
where $\kappa/2m_q$ ($\tilde{\kappa}/2m_q$) is the CMDM (CEDM) of the quark
$q$.  The Feynman rules for the interactions of quarks and gluons can be
written down:
\begin{equation}
\label{qqg}
{\cal L}_{q_i q_j g} = -g_s \bar q_j T^a_{ji} \left[ \gamma^\mu +
\frac{i}{2m_q} \sigma^{\mu\nu} p_\nu (\kappa - i \tilde{\kappa} \gamma^5) 
\right ] q_i\; G_\mu^a \;,
\end{equation}
where $q_i (q_j)$ is the incoming (outgoing) quark 
and $p_\nu$ is the 4-momentum
of the outgoing gluon.  The Lagrangian in Eq.~(\ref{eff}) also induces a 
$qqgg$ interaction given by 
\begin{equation}
\label{qqgg}
{\cal L}_{q_i q_jgg} = \frac{ig_s^2}{4m_q} \bar q_j (T^b T^c - T^c T^b)_{ji}
\sigma^{\mu\nu} ( \kappa -i \tilde{\kappa} \gamma^5) q_i G_\mu^b G_\nu^c \;,
\end{equation}
which is absent in the SM.  In the following, we write 
\begin{equation}
\kappa' = \frac{\kappa}{2m_q}\;, \tilde{\kappa}' = \frac{\tilde{\kappa}}{2m_q}
\end{equation}
which are given in units of (GeV)$^{-1}$.

\subsection{Prompt Photon Production}

\begin{figure}[ht]
\leavevmode
\begin{center}
\includegraphics[height=3in]{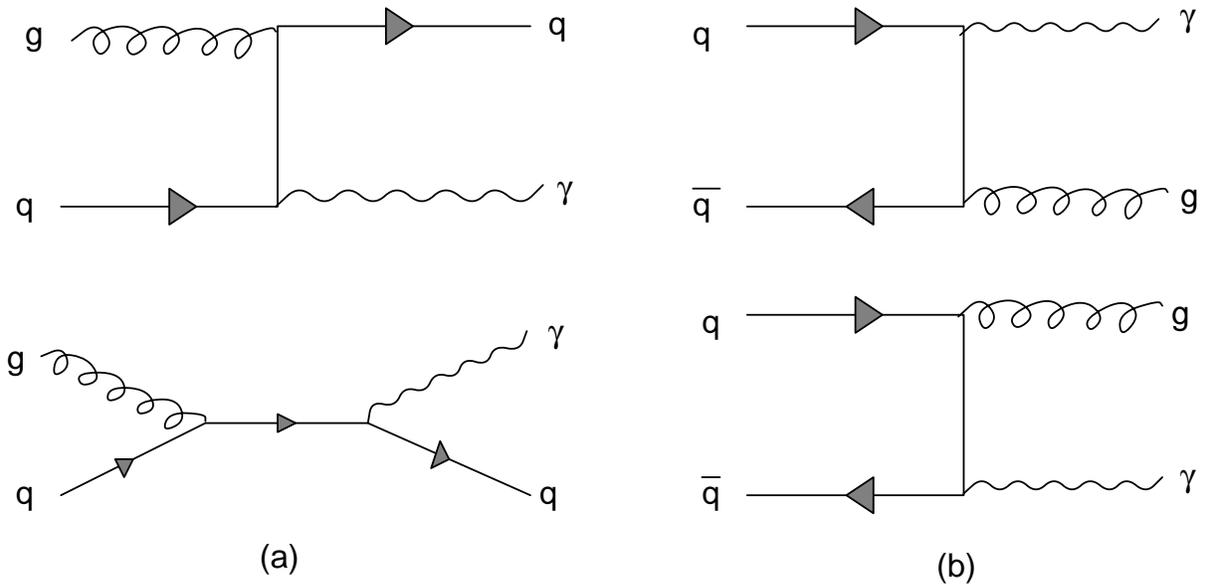}
\end{center}
\caption{Contributing Feynman diagrams for the processes: (a) $qg \to \gamma 
q$, and (b) $q\bar q \to \gamma g$.}
\label{fig1}
\end{figure}

 The contributing processes are:
\begin{displaymath}
q(\bar q) g \to \gamma q(\bar q) \;, \;\;\;\; q\bar q \to \gamma g \;.
\end{displaymath}
The contributing Feynman diagrams are shown in Fig.~\ref{fig1}.
The spin- and color-averaged amplitude for $q(p_1) g(p_2) \to \gamma(k_1) 
q(k_2)$ is given by
\begin{equation}
\label{amp1}
\overline{\sum} |{\cal M}|^2 = \frac{16\pi^2 \alpha_s
\alpha_{\rm em} Q_q^2}{3} \biggr [ - \frac{s^2 + t^2}{st} - 2 u 
(\kappa^{'2} + \tilde{\kappa}^{'2} ) \biggr]
\end{equation}
where
\begin{equation}
s=(p_1+p_2)^2\,, \;\; t=(p_1-k_1)^2 \,, \;\; u=(p_1-k_2)^2 \,,
\end{equation}
and $Q_q$ is the electric charge of the quark $q$ in units of the proton 
charge.
Similarly, the spin- and color-averaged amplitude for 
$q(p_1) \bar q (p_2) \to \gamma(k_1) g(k_2)$ is given by
\begin{equation}
\label{amp2}
\overline{\sum} |{\cal M}|^2 = \frac{128\pi^2 \alpha_s
\alpha_{\rm em} Q_q^2}{9} \biggr [ \frac{t^2 + u^2}{ut} + 2 s 
(\kappa^{'2} + \tilde{\kappa}^{'2} ) \biggr] \;.
\end{equation}
The subprocess cross section is then given by
\begin{equation}
d\hat \sigma = \frac{1}{(2\pi)^2 2 s} \overline{\sum} |{\cal M}|^2
\, \delta^4(p_1+p_2 - k_1 -k_2) \frac{d^3k_1}{2k_1^0}
\frac{d^3k_2}{2k_2^0}
\end{equation}
which is then folded with the appropriate parton distribution functions.
We use the CTEQ2M parton distribution functions\cite{cteq} 
and the two-loop formula for
the strong coupling constant.  Although the next-to-leading order (NLO)
 calculation to prompt photon production exists, there is, however,
 no NLO calculation that includes  CMDM and CEDM couplings.  
Therefore, throughout the paper we employ only
the leading order (LO) calculation.  But in calculating the fractional 
difference from the 
pure QCD cross section we shall use a K-factor to multiply the LO QCD
cross sections by.  The procedures will be illustrated in the next section.

\section{Results}

\begin{figure}[ht]
\leavevmode
\begin{center}
\includegraphics[height=5in]{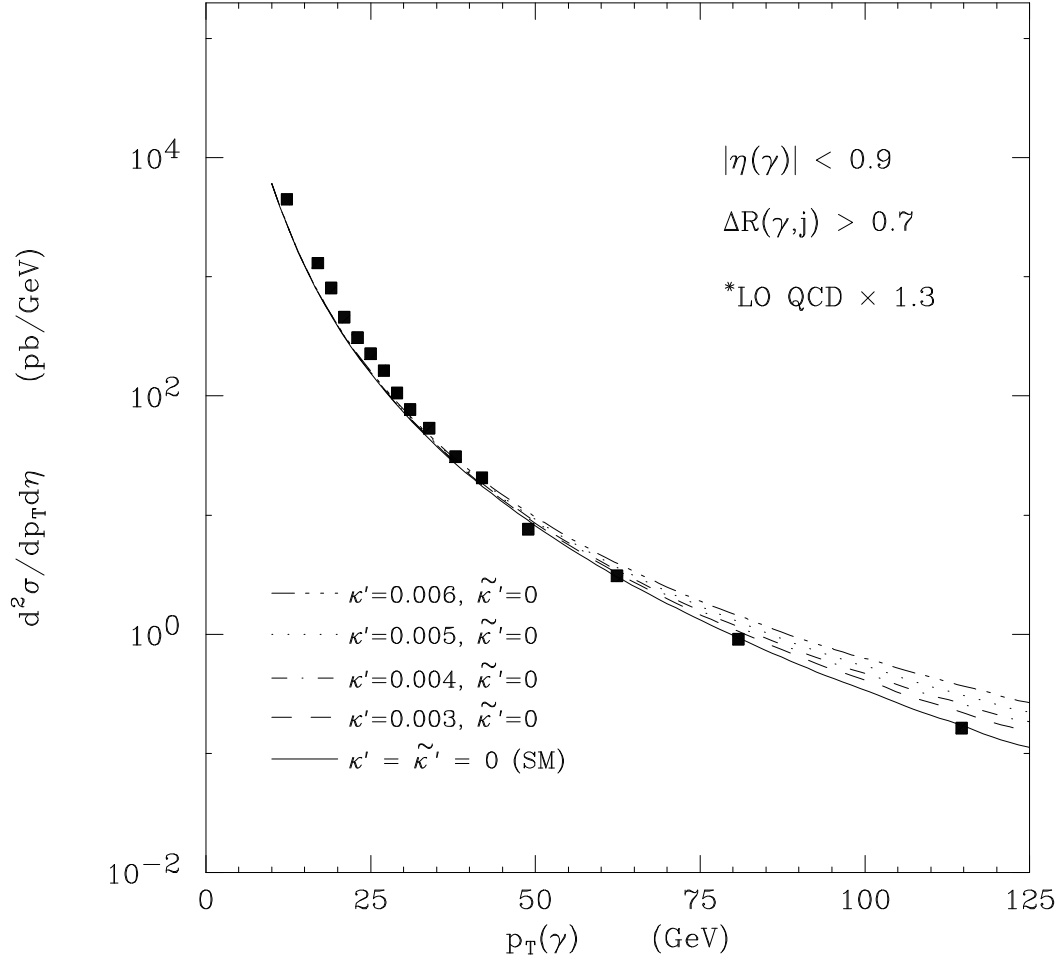}
\end{center}
\caption{Differential cross sections for prompt photon production versus 
the transverse momentum of the photon for pure QCD and nonzero values of 
CMDM of quarks. The data points are from CDF.}
\label{fig2}
\end{figure}

We first study the effects of nonzero CMDM and CEDM on the transverse 
momentum spectrum of the photon.  In order to compare with experimental 
data we have to impose a similar set of acceptance cuts as CDF and D0 did.
For both  CDF and D0 data we use
\begin{equation}
|\eta(\gamma)|<0.9\,, \;\; \Delta R(\gamma,j)>0.7 \,,
\end{equation}
where the $\Delta R(\gamma,j)$ cut is used to imitate the complicated 
experimental isolation procedures.  In our LO calculation, the value of this
$\Delta R(\gamma,j)$ cut is not crucial to our analysis. 
We have included the quark flavors: $u,d,s,c$ in our calculation and 
assumed that their anomalous couplings are the same.
In Fig.~\ref{fig2}, we show the differential cross sections of prompt
photon production versus the transverse momentum of the photon.  
The LO QCD curve has to be multiplied by a factor of about 1.3 to best fit the 
CDF data.  Therefore, we shall use a K-factor $K=1.3$ for the LO QCD 
cross section.   Figure \ref{fig2} also shows curves with nonzero values
of CMDM.  We can see that nonzero $\kappa'$ 
will increase the total and the differential cross sections, especially, 
in the large $p_T(\gamma)$ region.  Thus, the transverse momentum spectrum
becomes harder with nonzero CMDM.  
The effects due to nonzero CEDM will be the same because the increase in
cross sections is proportional to $(\kappa^{'2} + \tilde{\kappa}^{'2})$.
This is different from the case of $t\bar t$ production \cite{cheung}, 
in which the increase has a term proportional to the first power of $\kappa$.  

\begin{figure}[ht]
\leavevmode
\begin{center}
\includegraphics[height=5in]{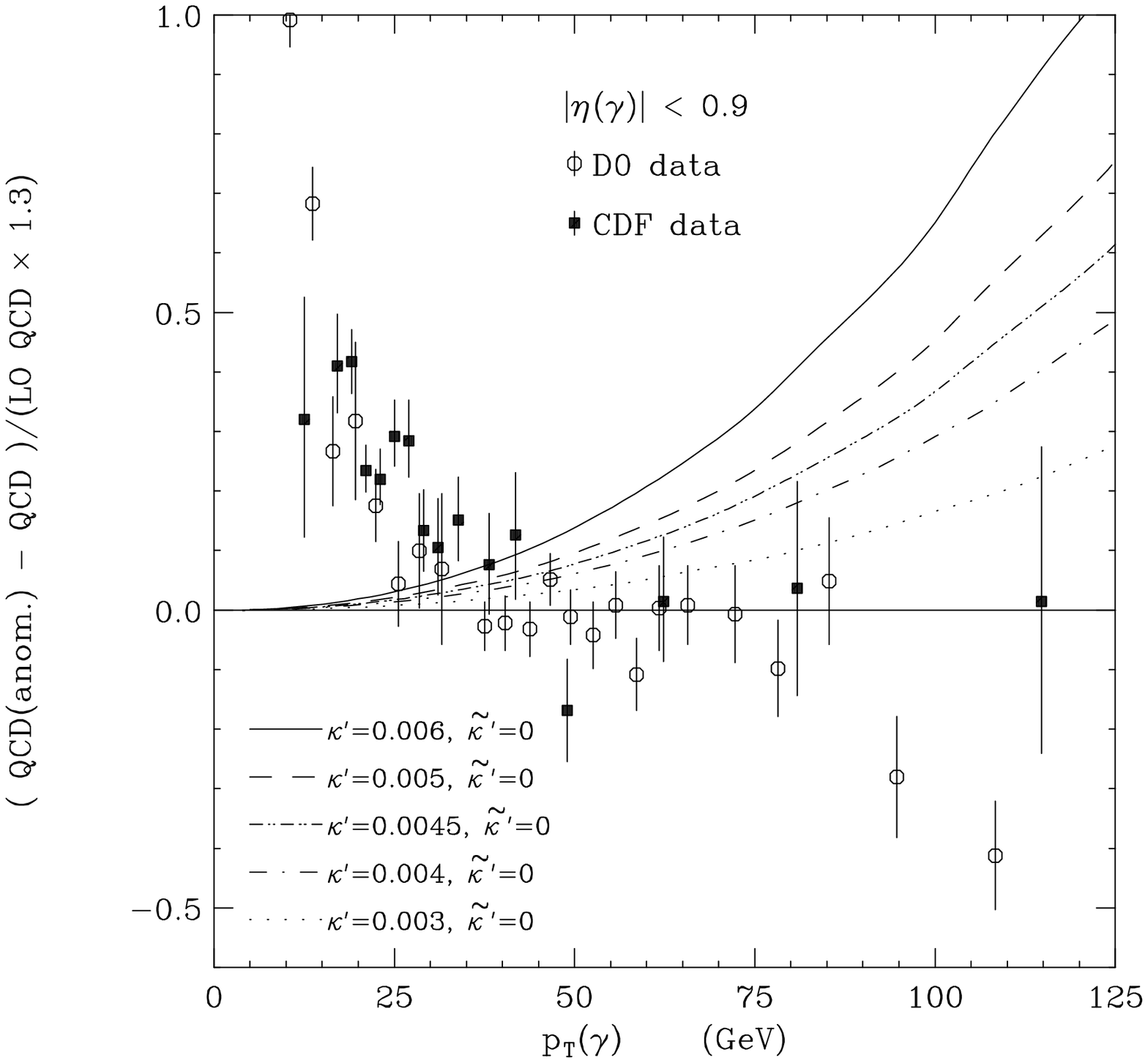}
\end{center}
\caption{Fractional difference from QCD for various values of $\kappa'$ and
$\tilde{\kappa}'$. 
Both D0 and CDF data are shown.  The
errors on D0 data are statistical only.}
\label{fig3}
\end{figure}

Figure \ref{fig3} shows the fractional differences from pure QCD for nonzero
CMDM.  The data are from CDF and D0.  The anomalous behavior at the
low $p_T(\gamma)$ has already been resolved by including initial and final
state shower radiation.  For our case we are only interested in the large
$p_T(\gamma)$ region.  Since in Eqs.~(\ref{amp1}) and (\ref{amp2}) the role of 
$\kappa'$ and $\tilde{\kappa}'$ are the same,  we can put one of them to be 
zero when we obtain bounds on the other.  We show a few curves with
different values of $\kappa'$.  From these curves we can see that the CDF 
and D0 data would be inconsistent with $\kappa' > 0.0045$, therefore, 
giving a bound of 
\begin{equation}
\kappa' \le 0.0045 \;\;{\rm GeV}^{-1}
\end{equation}
on the CMDM of quarks.  
Similarly, we put a bound of
\begin{equation}
\tilde{\kappa}' \le 0.0045 \;\;{\rm GeV}^{-1}
\end{equation}
on the CEDM of quarks.  
Furthermore, this bound is also valid for the case that the photon-quark
coupling is anomalous instead of the gluon-quark coupling.  
We also found that the normalized angular distribution in $\cos\theta^*$ 
($\theta^*$ is the angle of the outgoing photon in the CM frame of the
incoming partons) is not 
affected appreciably by the presence of the anomalous dipole moments.
We compare these with the results obtained in 
Ref.\cite{dennis}.  The value of $\kappa'$ obtained in fitting to the 
CDF\cite{cdf-jet} transverse energy distribution of the inclusive jet 
production without adjusting the gluon parton distribution function 
is \cite{dennis}
\begin{equation}
\kappa' = (1.0 \pm 0.3 )\times 10^{-3} \;\; {\rm GeV}^{-1}
\end{equation}
which is consistent with the bound obtained in this paper.

\section{Sensitivity at Future High Energy Collider Experiments}

The next run (Run II) at the Tevatron will be at $\sqrt{s}=2$ TeV with an 
integrated luminosity of $2\;{\rm fb}^{-1}$.  If the Run II is stretched
to a longer run it could accumulate a luminosity of about $10\;{\rm fb}^{-1}$
\cite{tev2000}.
There is also a plan called TeV33 \cite{tev2000} 
after Run II, in which the luminosity gets a 
further boost to about $30\;{\rm fb}^{-1}$. At about the same time scale the
CERN Large Hadron Collider (LHC) will operate at $\sqrt{s}=14$ TeV with
an initial yearly luminosity of $10\;{\rm fb}^{-1}$, which will later 
increase to the designed luminosity of $100\;{\rm fb}^{-1}$.
In this section, we shall estimate the sensitivities of $\kappa' \equiv
1/\Lambda$ or the limits on $\Lambda$ that can be probed in these future 
experiments.   We shall use a simple approach to calculate the limits.

Without a full Monte Carlo of the detector including energy
determination errors, we treat here only the statistical
sensitivity of the various experiments.  Our criteria\cite{hinchliffe}
is to take bins of appropriate size for the energy range being
examined, and find the $p_T$ called $p_T^*$ at which the SM cross
section statistical error bars are 10\%.  These will be the bins with
100 SM events.  We then explore the cross section due to
the SM plus the anomalous chromomagnetic moment contribution, and find
the value of $\kappa' \equiv 1/\Lambda$ or $\Lambda$ where the excess
over the SM is 10\% at this $p_T^*$.  In Table I we show the $p_T^*$ and 
$\Lambda \equiv 1/\kappa'$ for various experiments \cite{snow}.
We have used only the leading order cross sections without a $K$ factor
to determine $p_T^*$ and $\Lambda$.  
Since the $p_T$ distribution is steeply falling, 
so with or without a $K$ factor would not affect significantly 
the values for $p_T^*$ and $\Lambda$.
Actually, the experimental determination of $p_T$ might be the largest
systematic errors among all \cite{cdf-jet,cdf,d0}.
We have also imposed cuts on the isolated photon by 
$|\eta|<0.9$ and $\Delta R(\gamma, j)>0.7$ at the Tevatron energies, while
$|\eta|<1$ and $\Delta R(\gamma,j)>0.7$ for the LHC.
We can see from the table that $\Lambda$ sensitivity scales roughly with
the machine energy, but scales with about the eighth root of the luminosity.

\section{Conclusions}

We have studied the effects of anomalous chromomagnetic and chromoelectric
dipole moment couplings of light quarks on prompt photon production.  
The increase in cross sections is proportional to $\kappa^{'2} + \tilde
\kappa^{'2}$.  These couplings increase the total cross section and the 
transverse momentum spectrum, especially at the large $p_T(\gamma)$ region.
Using the CDF and D0 data we found a bound $\kappa'$ or $\tilde\kappa'
\le 0.0045\;{\rm GeV}^{-1}$ on the chromomagnetic or chromoelectric
dipole moment of light quarks, which is the main result of the paper.
In addition, we have also estimated the sensitivity of $\kappa'\equiv
1/\Lambda$ in the future collider experiments at the Tevatron and LHC.
The sensitivity is shown to be scaled roughly with the machine energy,
but with the eighth root of the luminosity.
For example, the Run II at the Tevatron can probe $\Lambda\equiv 1/\kappa'$
in the range $1.5 - 2$ TeV for integrated luminosities of 
$2 - 30\;{\rm fb}^{-1}$, while the LHC can probe up to 6 TeV with
a $100\;{\rm fb}^{-1}$ luminosity.

We acknowledge the supported from  U.S. DOE under grant 
No. DE-FG03-91ER40679 and No. DE-FG03-93ER40757

%

\vspace{1in}

\begin{table}[th]
\begin{center}
\caption{Table of high $p_T(\gamma)$ bins at 10\% statistical error and
1-$\sigma$ sensitivity for $\Lambda$ in that bin.}
\label{table1}
\begin{tabular}{|l||r|r|r||r|r|}
\hline
 & & Integrated & & \multicolumn{2}{c|}{Photons} \\
\cline{5-6} 
Accelerator & $E_{\rm cm}$ & Luminosity & Bin Width
& $p_T^*$ & $\Lambda$ \\
\hline
 & TeV & fb$^{-1}$ & GeV &  GeV & TeV \\
\hline
Tevatron:& & & & &  \\
         Run I  & 1.8 & 0.1 & 10 &  140 & 0.7 \\
         Run II & 2.0 & 2   & 20 &  260 & 1.5 \\
        Stretch & 2.0 & 10  & 20 &  325 & 1.9 \\
        TeV33   & 2.0 & 30  & 20 &  370 & 2.1 \\
\hline
        LHC     & 14  & 10  & 100 & 1000 & 4.5 \\
        LHC     & 14  & 100 & 100 & 1400 & 6.3 \\
\hline
\end{tabular}
\end{center}
\end{table} 

\begin{thebibliography}{2}
  
\bibitem{lep}P. B. Renton, OUNP-95-20, invited talk at LP'95: 
International Symposium on Lepton Photon Interactions (IHEP), Beijing, 
P.R. China, 10-15 Aug 1995. 

\bibitem{cdf-jet}CDF Collaboration (F. Abe {\it et al.}), 
Phys. Rev. Lett. {\bf 77}, 438 (1996).

\bibitem{weinberg}S. Weinberg, Phys. Rev. Lett. {\bf 37}, 657 (1976); 
Phys. Rev. Lett. {\bf 63}, 2333 (1989); Phys. Rev. {\bf D42}, 860 (1990).

\bibitem{cheung}K. Cheung, Phys. Rev. {\bf D53}, 3604 (1996).

\bibitem{rizzo}D. Atwood, A. Kagan, and T. Rizzo, Phys. Rev. {\bf D52}, 6254
(1995).

\bibitem{haberl}P. Haberl, O. Nachtmann, and A. Wilch, Phys. Rev. {\bf D53},
4875 (1996).

\bibitem{dennis}D. Silverman, hep-ph/9605318, to be published in Phys. Rev.
{\bf D}.

\bibitem{cdf}CDF Collaboration (F. Abe {\it et al.}), Phys. Rev. Lett. 
{\bf 73}, 2662 (1994).

\bibitem{d0}D0 Collaboration (S. Abachi {\it et al.}), FERMILAB-PUB-96/072-E.

\bibitem{cteq}CTEQ Collaboration (J. Botts {\it et al.}), Phys. Lett. {\bf 
B304}, 159 (1993).

\bibitem{tev2000}{\it Report of the Tev-2000 Study}, FERMILAB-Pub-96/082,
edited by D. Amidei and R. Brock.

\bibitem{hinchliffe} This is in the spirit of the U.S. ATLAS and U.S. 
CMS Coll., edited by
I. Hinchliffe and J. Womersley, LBNL-38997 (1996).

\bibitem{snow}A similar table for $\Lambda$ from high transverse energy jet
production has been presented in 
K. Cheung and D. Silverman, in {\it Proceedings of the 1996 DPF/DPB Summer 
Study on New Direction for High Energy Physics}, Snowmass CO, 1996 
(hep-ph/9609454).


\end{thebibliography}
\end{document}